\documentclass[
superscriptaddress,
preprint,
 amsmath,amssymb,
onecolumn,
]{revtex4-1}

\usepackage{graphicx}% Include figure files
\usepackage{dcolumn}% Align table columns on decimal point
\usepackage{bm}% bold math
\usepackage{lipsum}
\usepackage{color, soul}
\usepackage{enumitem}
\usepackage{ulem}
\begin{document}

\title{Removing proteins or bacteria on a tilted surface using air bubbles}% Force line breaks with \\

\author{Alireza Hooshanginejad}
 \affiliation{%
 Department of Biological and Environmental Engineering, Cornell University, Ithaca, New York, USA.
}%%Lines break automatically or can be forced with \\
 \affiliation{%
 School of Engineering, Brown University, Providence, Rhode Island, USA.
}%%Lines break automatically or can be forced with \\

\author{Timothy Sheppard}
 \affiliation{%
 Department of Biological and Environmental Engineering, Cornell University, Ithaca, New York, USA.
}%%Lines break automatically or can be forced with \\
\author{Purui Xu}
 \affiliation{%
 Department of Biomedical Engineering, Cornell University, Ithaca, NY, USA.
}%%Lines break automatically or can be forced with \\

\author{Janeth Manyalla}
 \affiliation{%
 Department of Biological and Environmental Engineering, Cornell University, Ithaca, New York, USA.
}%%Lines break automatically or can be forced with \\

\author{John Jaicks}
 \affiliation{%
 Department of Biological and Environmental Engineering, Cornell University, Ithaca, New York, USA.
}%%Lines break automatically or can be forced with \\

\author{Ehsan Esmaili}
 \affiliation{%
 Department of Mechanical Engineering, Purdue University, West Lafayette, Indiana, USA.
}%%Lines break automatically or can be forced with \\

\author{Sunghwan Jung}%
 \email{sunnyjsh@cornell.edu}
\affiliation{%
Department of Biological and Environmental Engineering, Cornell University, Ithaca, New York, USA.
}%

\date{\today}

\begin{abstract}
Cleaning surfaces with bubbles has been a topic of discussion in recent years due to the growing interest in sustainable methods for cleaning. Specifically, a method of using air bubbles to sanitize agricultural produce has been proposed as an eco-friendly alternative to current methods. In this study, we conduct experiments to test the cleaning efficacy at different angles of inclination of a contaminated surface. We use two different types of surface coated with either a protein solution or a bacterial biofilm. Our experimental results indicate that bubbles exhibit the best cleaning efficacy at the surface angle of $\theta \approx 20^{\rm{o}}$ for polydisperse bubbles in the range of 0.3-2 mm and with an average radius of 0.6 mm in radius. To gain a better understanding of the underlying mechanism, we perform a numerical analysis of a single air bubble impacting surfaces with different angles. Our numerical and theoretical results show that the shear stress, which is proportional to the sliding speed but inversely proportional to the thickness of the film, results in the maximum shear force occurring at $\theta\approx22^{\rm{o}} \approx \pi/8$ which agrees well with the experiments. 
\end{abstract}
\maketitle

\section{Introduction}
Multiphase flows have been used for the removal of micrometer-scale contaminants from surfaces for decades \cite{OBrien1991,Gomez1999,khodaparast2017bubble}. Specifically, air bubbles are proposed as a sustainable cleaning method for wastewater treatments \cite{gogate2004review,chan2009review,temesgen2017micro}, or preventing biofouling \cite{menesses2017measuring,Hopkins2021}. Recently, agricultural produce such as fruits and vegetables has been cleaned with the insertion of air bubbles \cite{Sunny1,lee2018cavitation,Esmaili2019}. Developing such environmentally benign methods for sanitizing agricultural produce is important, as pathogens in fresh produce are recognized as the primary cause of foodborne diseases in millions of people each year \cite{scallan2011foodborne,scallan2011foodborne2,morris2011safe}. From a practical point of view, numerical analysis of a bubble impacting and sliding on a tilted surface has shown that the maximum shear stress exerted on the surface is sufficient for removing different types of bacteria from the surface \cite{Esmaili2019}.

Bubble dynamics has been studied extensively in both a freely rising bubble \cite{clift2005bubbles,duineveld1995rise,amaya2010single,peters2012experimental,wu2002experimental} and a bubble interacting with either horizontal \cite{manica2015force,zawala2016immortal,zawala2016influence,krasowska2007kinetics,Zenit2009,klaseboer2014force} or vertical \cite{Vries2002,takemura2003transverse,Moctezuma2005,FIGUEROA-ESPINOZA2008} solid surfaces. However, only a few studies have investigated the dynamics of bubbles along a tilted solid surface \cite{maxworthy1991bubble,tsao1997observations,Perron2006,podvin2008model,norman2005dynamics,debisschop2002bubble,Barbosa2019}. Specifically, by incorporating lubrication approximation the thin film flow between the sliding bubble and the surface can be modeled \cite{podvin2008model,Esmaili2019}. Then the shear stress can be quantified on the surface, which is the key in characterizing the cleaning effect of sliding bubbles. Although shear stress was recently calculated for varying surface angles \cite{Esmaili2019}, no experiments have yet verified the cleaning effect of air bubbles on tilted surfaces. In addition, while the maximum shear stress on the surface has been reported for different inclination angles \cite{Esmaili2019}, the average shear stress has not been studied for practical applications and comparison with potential experiments. 

In this study, we perform experiments to use bubbles of millimetric scale and clean surfaces coated with either protein or bacterial coatings at different tilting angles of the surface. Our results show that there exists a critical angle for maximum cleaning effect. In addition, we perform numerical calculations of an air bubble impacting and sliding over the tilted surface incorporating the recent model on bubble dynamics near a tilted wall \cite{Esmaili2019}. The manuscript is organized as follows. In section \ref{sec:exp}, we discuss experimental methods including the coating procedure in section \ref{subsec:coating_methods} and the cleaning experiments in section \ref{subsec:cleaning_exp}, and numerical methods in section \ref{subsec:numerical_method}. We then present and discuss our experimental and computational results in section \ref{subsec:exp_result}. Finally, in section \ref{sec:summary} we summarize our findings and discuss future studies.

\section{Materials and Methods} \label{sec:exp}
% In this section we discuss our experimental procedure for investigating cleaning effect of air bubbles impacting surfaces with different inclination angles. We first explain the coating process for both a protein solution, and a bacteria biofilm in section \ref{subsec:coating_methods}. We then explain the bubble experiments in section \ref{subsec:cleaning_exp}.

\subsection{Coating Preparation}  \label{subsec:coating_methods}
\subsubsection{Protein Dirt Preparation} \label{subsubsec:protein_soil}
A protein dirt/soil solution is synthesized for glass slide coating. We first incorporate 100 g of 2\% fat milk and 30 g of sifted wheat flour in a small pot with an immersion blender until the mixture reaches 115 $^{\rm{o}}{\rm{C}}$, and a paste-like solid is formed. This mixture is then left to cool to 30 $^{\rm{o}}{\rm{C}}$, wherein afterwards it is combined with another 120 g of milk and is blended again until a uniform solution remains after 10 minutes. Next, 6 g of Nigrosin dye is gently stirred in with a wooden tongue depressor until fully combined. Once completed, the mixture is sifted twice, first through a 500-600 $\mu \rm{m}$ mesh, then again through a 100-150 $\mu \rm{m}$ mesh. The resulting mixture is then ready to be used for coating.

\subsubsection{Spin Coating of Protein Dirt}  \label{subsubsec:spin_coater}
On the same day as protein preparation, glass microscope slides of 76.2 mm $\times$ 25.4 mm $\times$ 1.1 mm are cleaned and coated with the protein solution using a custom-designed spin coater. More information about making the spin coater including 3D printing designs can be found in \cite{Hooshanginejad2022}. To coat the slides, we first secure the slide tightly to the spin-coater plate with opposing metal bolts. Next, a 0.5 mL protein drop is gently deposited onto the middle of the slide and spinned for 10 seconds at a rate of $\Omega\sim 1100-1200$ rotations per minute (RPM). These coated slides are left to dry for 15 minutes inside a fume hood, then proceeded by a second coating of the same procedure. This second coating creates a more opaque surface which proved to be more effective in characterizing the cleaning effect of our tests. After the second coating, the slides are stored in a cool, dry place. In the current study, all slides are tested after drying for 2 full days. We note that the effect of the drying time was investigated in previous studies showing a decaying cleaning effect as the coating dries for longer periods \cite{Hooshanginejad2022}.
 
 \subsubsection{Bacterial Coating Preparation} \label{subsubsec:bacterial coating}
To test the bubble-cleaning effect with live organisms, we also prepare glass slides coated with \textit{E. coli}. For better visualization, we introduce the green fluorescent protein (GFP) gene into \textit{E. coli} (MM294 strain, Carolina Biological Supply Co.) with the heat-shock plasmid transformation method. Using a disposable sterile pipette, we add 250 $\mu$L of 50 mM $\rm{CaCl_{2}}$ solution to a sterile micro-tube. Then we use a sterile inoculation loop to transfer 5 \textit{E.coli} colonies from the source media to the tube, and immerse the loop tip into the $\rm{CaCl_{2}}$ solution and vigorously span the loop to disperse the entire mass into the $\rm{CaCl_{2}}$ solution. The tube is then placed in an ice bath. Next, we transfer 10 $\mu$L of 0.01 $\mu$g/$\mu$L pGREEN (4528 bp) directly into the tube using a sterile pipette, and gently mix the solution with the pipette. The tube is incubated in an ice bath for 10 minutes. After that, we take the tube to the 42$^{\rm{o}}{\rm{C}}$ water bath, and hold the tube under water for 45 seconds to give the bacteria colony a heat shock, and immediately return the tube to ice. After 2 minutes, we add 250 $\mu$L of recovery Luria broth (LB) to the tube. We gently mix the solution, and incubate the tube at room temperature for 10 minutes. We label 1 LB agar plate and 1 LB/Ampicillin+GFP agar plate. The Ampicillin plate is critical for selecting the GFP transformed bacteria colonies, and is made with 100 $\mu$L of 10 mg/mL Ampicillin solution and 20 mL agar media. Using a sterile pipette, we transfer 50 $\mu$L of bacteria suspension from the tube to each plate and spread the bacteria. The plates are rested for 10 minutes, then sealed and incubated at 37$^{\rm{o}}{\rm{C}}$ upside down. After 24 hours of incubation, the GFP transformed colonies start to appear on the LB/Ampicillin+GFP plate. Using a sterile inoculation loop, we transfer the GFP transformed colonies to the LB broth and make a GFP \textit{E. coli} suspension.

Next, in an aseptic experimental environment, we juxtapose four identical sterile glass slides inside a sterile Petri dish (140 mm diameter). We then add 33 mL of liquid LB agar media into the Petri dish to fully cover the glass slides by agar media. We let the agar cool down and solidify. The resulting thickness of the agar on each slide is 1 mm. We use a sterile pipette and transfer 25 $\mu$L of living \textit{E. coli} suspension to the middle of the agar plate. We then roll sterile glass beads on the plate to evenly distribute the bacteria. Afterwards, the Petri dish is sealed with parafilm, and is rested for 10 minutes to let the agar plate absorb the bacteria suspension. Finally, the Petri dish is incubated at 37$^{\rm{o}}{\rm{C}}$ with the agar side on top for 72 hours. After the incubation, we gently cut the slides with the bacteria-agar layer on the top to use for the experiments.

\begin{figure}[t]
\includegraphics[width=\textwidth]{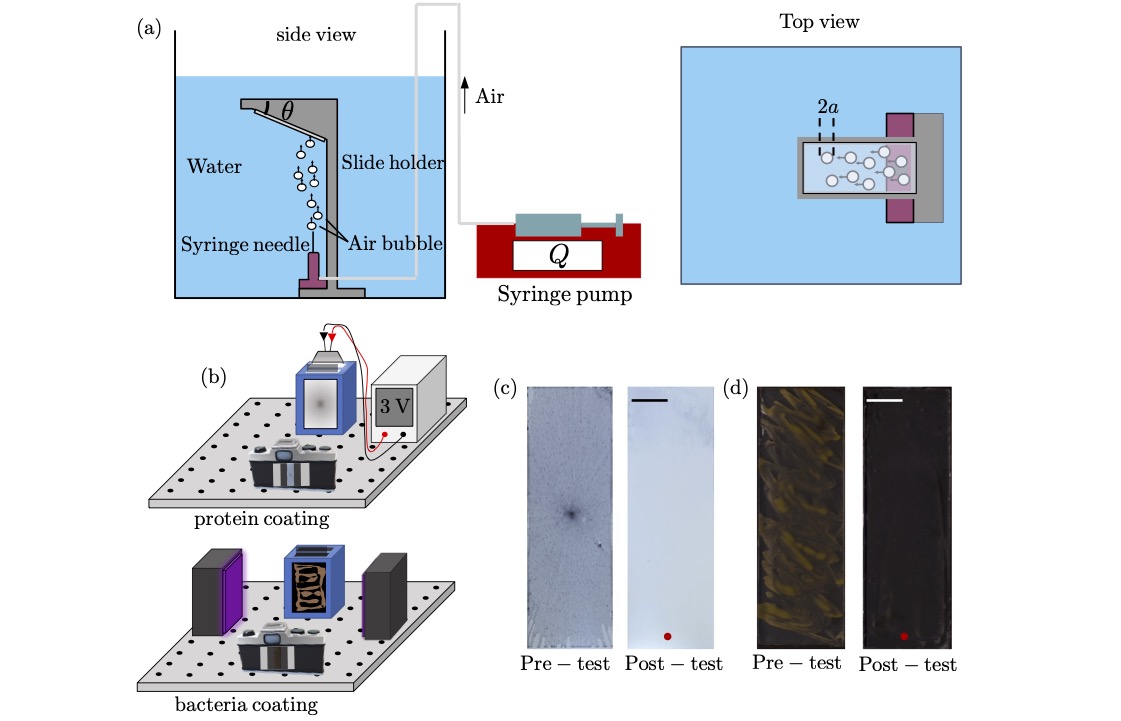}
\caption{\label{fig:exp_schematic} (a) side-view and top-view schematics of the experiments. (b) Schematics of the slide imaging setup. (c) Pre-test and post-test images of a protein-coated slide for $\theta=20^{\rm{o}}$ and 6 minutes of bubble testing. (d) Pre-test and post-test images of a biofilm coated slide for $\theta=20^{\rm{o}}$ and 6 minutes of testing. The red dots in the post-test images show the location of the bubble releaser. }
\end{figure}

\subsection{Bubble Cleaning Experiment} \label{subsec:cleaning_exp}
A 20 L tank is first filled with room temperature deionized water. Within it, a 3D printed slide-carrying tower holds coated slides at an angle, $\theta$, with regards to the bottom of the tank as shown in Fig. \ref{fig:exp_schematic}(a). At the base of this slide carrier, a 25 gauge syringe needle is placed 5 mm from the slide edge at a constant height of 11 cm from the surface. A 3D printed needle holder connects the syringe needle to an external syringe pump. Figure \ref{fig:exp_schematic}(a) shows the schematics for the side view and top view of the experiments.

Slides coated with proteins are imaged before and after each cleaning test utilizing a 3D printed LED slide stand fixed to an optical breadboard as illustrated in Fig. \ref{fig:exp_schematic}(b). The LED is maintained at 3V using a DC power supplier. The LED is replaced with a uniform black-painted background illuminated by a set of UV lights for the bacteria-coated slides. A digital camera (Nikon 7500) is fixed to the board 1.5 ft away from the slide with a 105 mm macro lens. The system is utilized in a dark room. Slide images from before and after the tests are then analyzed by performing image processing. We note that all the post-test images were taken after one hour of drying in room temperature. Figures \ref{fig:exp_schematic}(c) and \ref{fig:exp_schematic}(d) show sample pre-test and post-test images with protein coating and bacteria coating, respectively. We note that for protein coatings, there is always a darker spot appearing in the center of the slide indicative of a thicker coating layer, while some areas near the edges are thinner (see Fig. \ref{fig:exp_schematic}(c)). However, since all slides are coated with the same method and include these features, the effect of such spots on the average cleaning results is negligible. On the contrary, surfaces coated with bacteria indicate more variations in the thickness of the biofilm, as shown in the pretest image of Fig. \ref{fig:exp_schematic}(d). We discuss the effect of such randomness in biofilm thickness on the cleaning results in section \ref{sec:summary}. To characterize the efficacy of bubble cleaning, the intensity matrix of the pre-test, post-test, and clean-slide images are converted to grayscale matrices of $I_1$, $I_2$, and $I_0$, respectively. Hence, we define the total cleaning efficacy as $\lambda^{\rm{t}}=(I_2-I_1)/(I_1-I_0)$. Therefore, as $\lambda^{\rm{t}}$ approaches 1, all coated contaminants are removed from the surface. To isolate the role of bubbles in cleaning, we run a series of control experiments with the same $\theta$ and $T$ for each case where no bubbles are injected. We define the cleaning efficacy of the control experiments without bubbles in a similar way and denote it by $\lambda^{\rm{c}}$. Then, we define the efficacy of cleaning bubbles as $\lambda=\lambda^{\rm{t}}-\lambda^{\rm{c}}$. It is noteworthy that $\lambda^{\rm{c}}$ for all cases is quite small compared to $\lambda^{\rm{t}}$ and does not change the trend between $\lambda^{\rm{t}}$ and $\lambda$ significantly.

\begin{figure}
\begin{center}
\includegraphics[width=\textwidth]{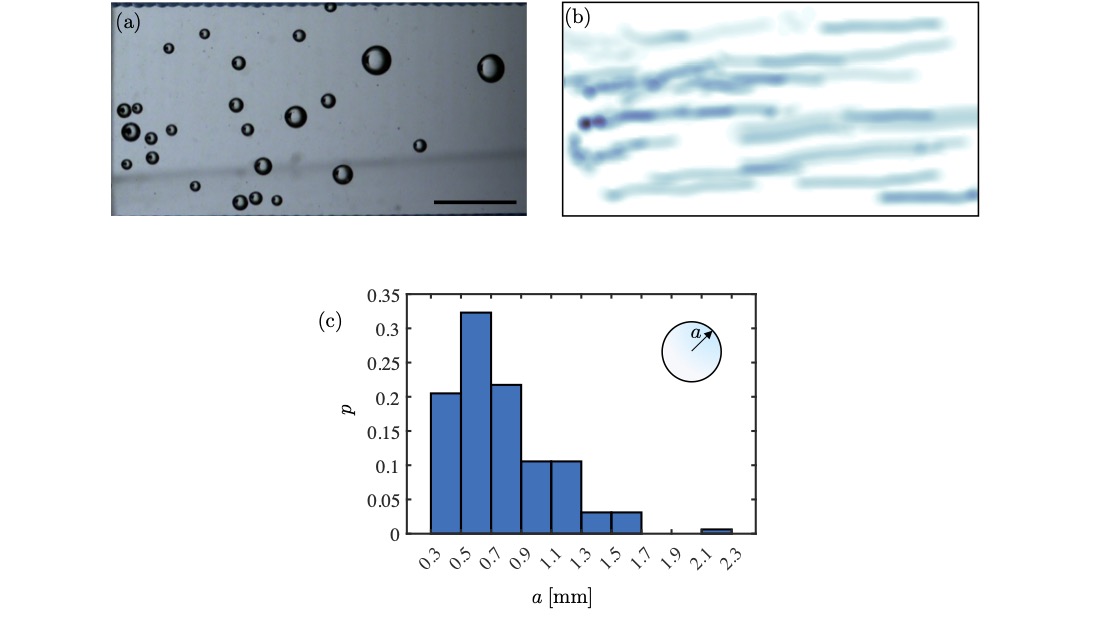}
\caption{\label{fig:histogram} (a) a snapshot of bubbles moving along a clean surface with $\theta=10^{\rm{o}}$. (b) trajectory of air bubbles for 0.4 seconds presented using an arbitrary color with a Gaussian blurring method at bubble centers. The Gaussian distribution has a root mean square width equal to a bubble radius. (c) Histogram of bubbles radius, $a$, for 5 different tests and approximately 200 bubbles. }
\end{center}
\end{figure}

\subsection{Numerical Methods} \label{subsec:numerical_method}
To gain a more fundamental understanding of $\lambda$, we implement a numerical model that includes all forces that contribute to the dynamics of a bubble impacting a tilted surface: the buoyancy force, the drag force, the lift force, the added mass force, and the thin film force. We expect the liquid film thickness to be on the order of micrometers and a stable liquid film force remains significant as the bubble moves near the tilted wall \cite{manica2015force,Esmaili2019}. This model has previously been discussed and studied for a similar condition where the bubble impacts and slides over a tilted wall \cite{Esmaili2019}. Hence, we only summarize the main features of the model herein and refer to \cite{Esmaili2019} for further details. We note that in this study we focus on features of bubble dynamics that are different from previous work \cite{Esmaili2019} to rationalize our current experimental results.

\begin{figure}
\includegraphics[width=\textwidth]{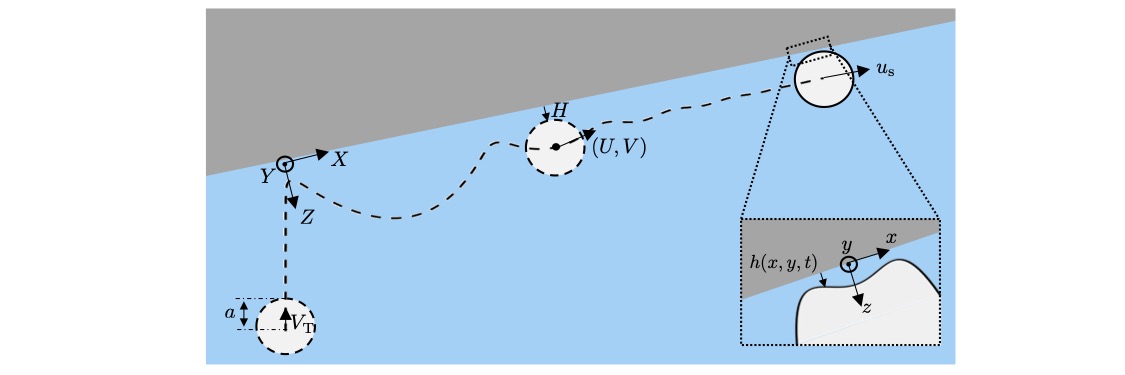}
\caption{\label{fig:num_schematic} Schematics of a bubble impacting a tilted surface from the bouncing to the sliding regime.}
\end{figure}

Figure \ref{fig:num_schematic} shows the schematic of a bubble impacting a tilted surface to introduce the notations used in our model. Both $XYZ$ and $xyz$ coordinates denote the axial direction, the transverse direction, and the direction normal to the surface, respectively. The $XYZ$ origin is located on the first impact point, while the $xyz$ origin is the bubble's centroid projected onto the surface which moves with the bubble. The bubble's centroid distance normal to the surface is denoted by $H$ while it moves with velocities $U$, and $V$, along $X$ and $Y$, respectively. In addition, the thin film thickness and the thin film pressure are denoted by $h(x,z,t)$ and $P(x,z,t)$, respectively. Hence, the force balance in $X$ and $Y$ yields as \cite{Esmaili2019}

\begin{multline}
\label{eqn:x-momentum}
\rho \Omega\left[C^{||}_{\rm{m}} \frac{dU}{dt}-\frac{d C^{||}_{\rm{m||}}}{dH}VU \right]\\
=\rho g \Omega \sin{(\theta)}-\frac{\pi}{4}C^{||}_{\rm{D}} {\rm{Re}}\mu Ua-\iint_{\rm{A}} P_{\rm{f}} \frac{dh}{dx}dx dy+ C_{\rm{L}}\rho\Omega V \omega_y,
\end{multline}
and,
\begin{multline}
\label{eqn:y-momentum}
\rho \Omega\left[C^{\bot}_{\rm{m}} \frac{dV}{dt}+\frac{1}{2}\left( \frac{d C^{||}_{\rm{m}}}{dH}U^2- \frac{d C^{\bot}_{\rm{m}}}{dH}V^2\right) \right]\\
=\rho g \Omega \cos{(\theta)}-\frac{\pi}{4}C^{\bot}_{\rm{D}} {\rm{Re}}\mu Va-\iint_{\rm{A}} P_{\rm{f}} dx dy- C_{\rm{L}}\rho\Omega U \omega_y,
\end{multline}
respectively. Here, $\rho$ denotes the air density, $\Omega$ denotes the bubble volume, and $\mu$ denotes the dynamic viscosity of water. Re=$2\rho_{\rm{w}}V_{\rm{T}}a/\mu$ is the Reynolds number with $\rho_{\rm{w}}$ denoting the water density and $V_{\rm{T}}$ denoting the bubble's terminal rising velocity. $C_{\rm{m}}$, $C_{\rm{D}}$, and $C_{\rm{L}}$ denote the coefficients of added mass, drag, and lift, respectively. In addition, the components of coefficients along $X$ and $Y$ are denoted with superscripts $||$ and $\bot$, respectively. The terms on the left hand side of Eq. (\ref{eqn:x-momentum}) and Eq. (\ref{eqn:y-momentum}) refer to the inertia terms including the added mass force, while the right-hand side terms correspond to the buoyancy force, the drag force, the thin film force, and the lift force, respectively. More details about the expressions used for each coefficient and how the thin film force is calculated can be found in \cite{Esmaili2019}.

The domain is divided into $105\times 105$ nodes. The equations are discretized using a finite difference method and are solved using a MATLAB ode15s solver. We make sure that all forces are continuously computed over the entire simulation time without any discontinuity. Taking into account the size distribution of the bubbles presented in Fig. \ref{fig:histogram}(c), we incorporate $a=0.6$ mm into our model, as it represents the most frequent bubble size in the experiments. It is assumed that the bubble is at $H=3$ mm from the surface at $t=0$ while rising with a measured terminal rising velocity, $V_{\rm{T}}\approx 32 $ cm/s. We model the bubble motion from $t=0$ until it reaches the point 6 cm downstream the first impact point along the solid surface. The 6 cm threshold value is chosen to match the average distance traveled by the bubbles in the current experiments. We note that the 6 cm threshold is well beyond the threshold of transitioning from bouncing to sliding regime. 

%a 1/169 shutter speed, an ISO of 1600, and F13 exposure
\section{Results} \label{subsec:exp_result}

\begin{figure}
\includegraphics[width=\textwidth]{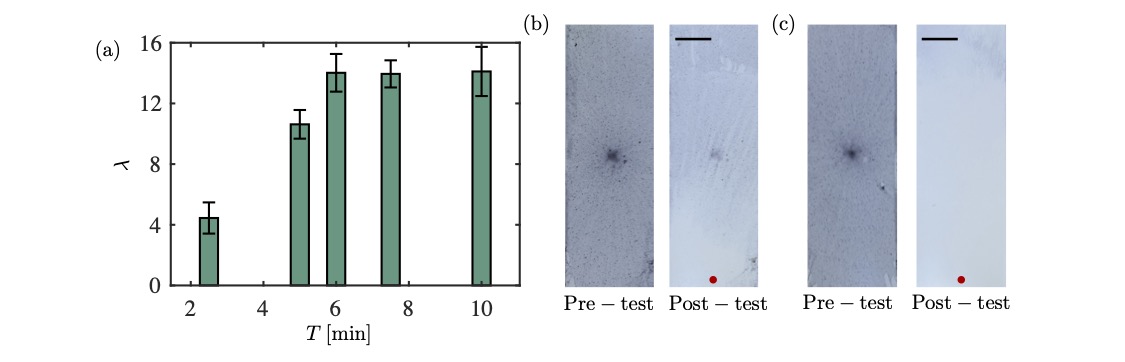}
\caption{\label{fig:T_lambda} (a) The cleaning efficacy, $\lambda$, for varying test time, $T$, at $\theta=20^{\rm{o}}$. (b) Pre-test and post-test images for $T=5$ min, and $\theta=20^{\rm{o}}$. The scale bar shows 1 cm. (c) Pre-test and post-test images for $T=6$ min, and $\theta=20^{\rm{o}}$. The red dots in post-test images show the location of the bubble releaser. The scale bar shows 1 cm.}
\end{figure}

We first extract the size distribution for bubbles generated from the 25 gauge needle used at a constant flow rate of $Q=10$ mL/min in all experiments. Figure \ref{fig:histogram}(a) shows a sample snapshot of bubbles moving along a surface with $\theta=10^{\rm{o}}$ from the top view. Figure \ref{fig:histogram}(b) shows surface areas covered by air bubbles over a duration of 0.4 s indicated by using a Gaussian blur at the centers of the bubbles with a root mean square width equal to their radii for the purpose of demonstration. As shown in Fig. \ref{fig:histogram}(b), bubbles adequately cover the slide along the transverse direction. The darker colors near the left end indicate the slower tangential motion of the bubbles along the surface near their impact point. Figure \ref{fig:histogram}(c) shows the probability histogram of the extracted bubble radius, $a$, for five different trials over 200 bubbles. Here, $a=0.6$ mm is the most frequent average bubble size as the bubble radius in simulations.

\begin{figure}
\begin{center}
\includegraphics[width=\textwidth]{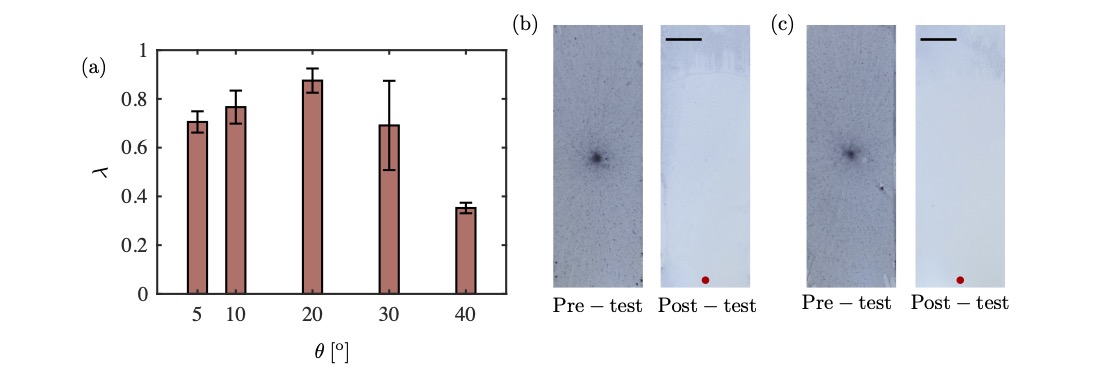}
\caption{\label{fig:protein_lambda} (a) The efficacy of cleaning, $\lambda$, on surfaces coated with proteins for varying surface angles, $\theta$, with $T=6$ min. (b) Pretest and post-test images for $\theta=10^{\rm{o}}$, and $T=6$ min. The scale bar shows 1 cm. (c) Pre-test and post-test images for $\theta=20^{\rm{o}}$, and $T=6$ min. The red dots in the post-test images show the location of the bubble releaser. The scale bar shows 1 cm.  }
\end{center}
\end{figure}

We first run a series of protein tests with different cleaning times, $T$. Figures \ref{fig:T_lambda}(b) and \ref{fig:T_lambda}(c) show the pre-test and post-test images at $\theta=20^{\rm{o}}$ with two different times: $T=5$ min and $T=6$ min, respectively. Figure \ref{fig:T_lambda}(a) shows $\lambda$ for $T=$2.5, 5, 6, 7.5, and 10 min with $\theta=20^{\rm{o}}$. As indicated in Fig. \ref{fig:T_lambda}(a), $\lambda$ increases with $T$ until it plateaus out after $T=6$ min. Therefore, we choose $T=6$ min as the reference test time on surfaces coated with proteins with different $\theta$. Next, we perform the cleaning experiments at five different angles, $\theta=5^{\rm{o}},\: 10^{\rm{o}}, \: 20^{\rm{o}},\: 30^{\rm{o}},$ and $40^{\rm{o}}$ for 6 minutes. As shown in Fig. \ref{fig:protein_lambda}(a), the cleaning efficacy, $\lambda$, first increases with angle up to $\theta>20^{\rm{o}}$, then decreases rapidly. Two representative cases ($\theta=10^{\rm{o}}$ and $\theta=20^{\rm{o}}$) are shown in Fig. \ref{fig:protein_lambda}(b) and Fig. \ref{fig:protein_lambda}(c), respectively.

\begin{figure}
\begin{center}
\includegraphics[width=\textwidth]{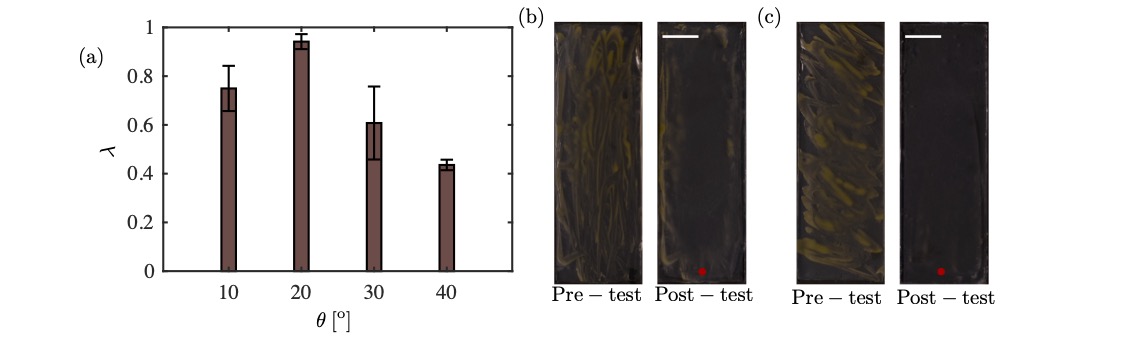}
\caption{\label{fig:biofilm_lambda}(a) The cleaning efficacy, $\lambda$, on surfaces coated with bacteria for varying surface angles, $\theta$, with $T=6$ min. (b) Pre-test and post-test images for $\theta=10^{\rm{o}}$, and $T=6$ min. The scale bar shows 1 cm. (c) Pre-test and post-test images for $\theta=20^{\rm{o}}$, and $T=6$ min. The red dots in post-test images show the syringe needle projection on the surface. The scale bar shows 1 cm.  }
\end{center}
\end{figure}

In addition, we conduct cleaning experiments with surfaces coated with bacterial biofilm following the procedure described in section \ref{subsubsec:bacterial coating}. We note that the biofilm tests are all conducted for 2 minutes to avoid any delamination of the film and catch up the fast removal time. Also, we notice that $\lambda$ does not increase noticeably for bacteria tests beyond 2 minutes of experiment. Figure \ref{fig:biofilm_lambda} (a) shows that the maximum $\lambda$ occurs at $\theta=20^{\rm{o}}$ similarly to the protein tests. Figures \ref{fig:biofilm_lambda}(b) and \ref{fig:biofilm_lambda}(c) show the pre-test and post-test images from two cases corresponding to $\theta=10^{\rm{o}}$ and $\theta=20^{\rm{o}}$, respectively. 

%\section{Theory} \label{sec:theory}

\begin{figure}
\includegraphics[width=\textwidth]{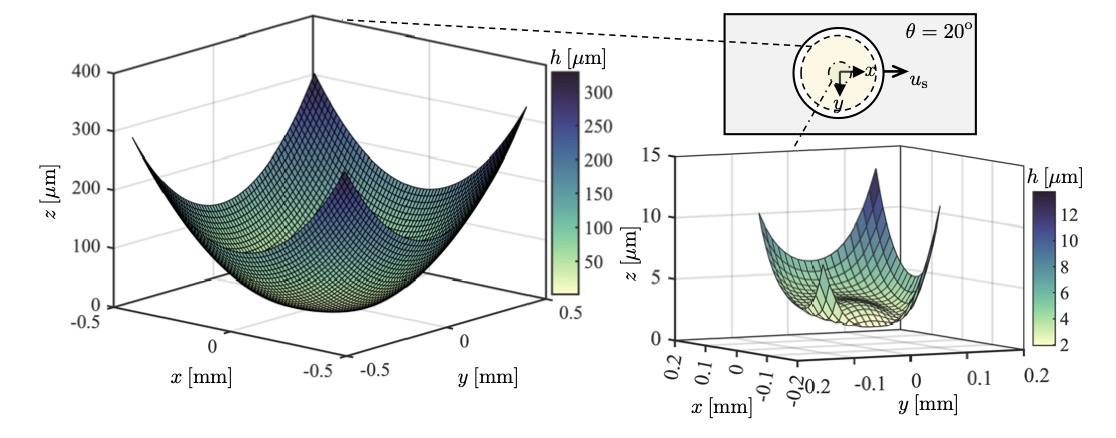}
\caption{\label{fig:3d_profile} Bubble profile for $a=0.6$ mm and $\theta=20^{\rm{o}}$ during the steady state sliding. The inset shows the zoom-in view of the dimple formed on the bubble, near the solid surface.}
\end{figure}

% \subsection{Results}  \label{subsec:numerical_results}
Next, let us discuss the results obtained from the numerical model that was described in section \ref{subsec:numerical_method}. The average shear force, $\overline{F}_{\rm{s}}$, exerted by the bubble on the surface is the key parameter relevant to cleaning experiments. To characterize $\overline{F}_{\rm{s}}$, we need to discuss two primary factors that affect the shear force of a bubble: the steady film thickness and the steady sliding speed. Figure \ref{fig:3d_profile} shows the 3D bubble shape during the steady sliding regime for $a=0.6$ mm, and $\theta=20^{\rm{o}}$. During the steady sliding regime, the bubble shape and the thin film profile do not change noticeably. Figure \ref{fig:3d_profile} inset shows the zoom-in view of the 3D bubble shape close to the surface where a dimple forms as previously reported \cite{platikanov1964experimental,pan2011effect,Esmaili2019}. In addition, Fig. \ref{fig:3d_profile} reveals a thinner neck forming on the receding side of the dimple compared to the advancing side. The film thickness on the receding side is of great importance, as the shear force on the surface is mainly composed of the cumulative shear force terms from the mesh points in this thin region. 

\begin{figure}
\includegraphics[width=\textwidth]{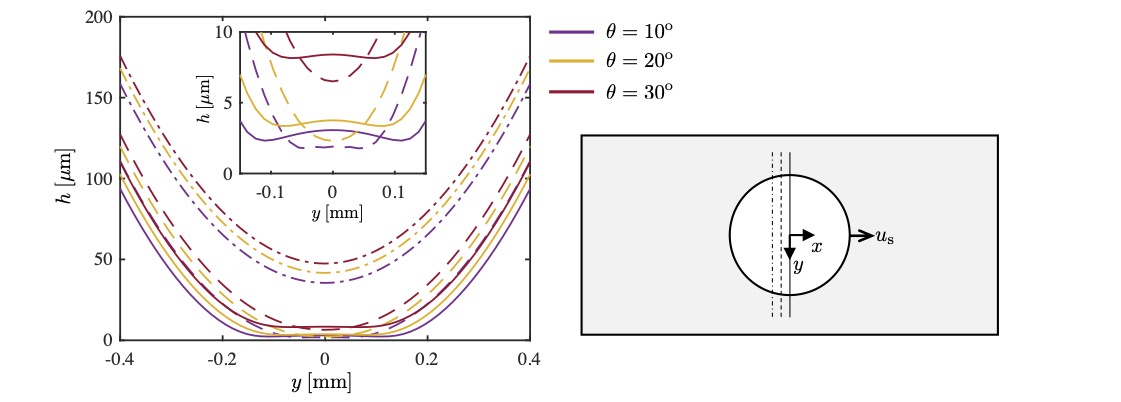}
\caption{\label{fig:x_profile} The bubble profile during the steady state sliding in $xz$ planes at $y=0$ $\mu$m (solid line), $y=\pm 100$ $\mu$m (dashed line), and $y=\pm 200$ $\mu m$ (dotted dashted line) with $a=0.6$ mm for $\theta=10^{\rm{o}}$ (violet), $\theta=20^{\rm{o}}$ (yellow), and $\theta=30^{\rm{o}}$ (red). The inset shows the zoom in view of the profile near the centerline. }
\end{figure}

\begin{figure}
\includegraphics[width=\textwidth]{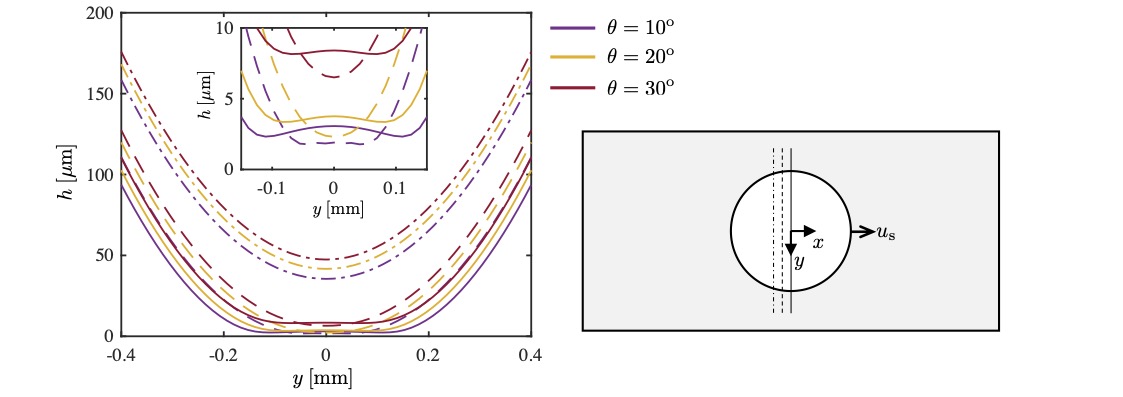}
\caption{\label{fig:y_profile} The bubble profile during the steady state sliding in $yz$ planes at $x=0$ $\mu$m (solid line), $x=\pm 100$ $\mu$m (dashed line), and $x=\pm 200$ $\mu m$ (dotted dashted line) with $a=0.6$ mm for $\theta=10^{\rm{o}}$ (violet), $\theta=20^{\rm{o}}$ (yellow) and $\theta=30^{\rm{o}}$ (red). The inset shows the zoom-in view of the profile near the centerline. }
\end{figure}

To gain better insight into the shear force on the surface, we explore the 2D bubble profile in the $xz$ plane for $y=0$ $\mu$m, 100 $\mu$m, and 200 $\mu$m. Figure \ref{fig:x_profile} indicates the bubble profile for $\theta=10^{\rm{o}}$, $\theta=20^{\rm{o}}$, and $\theta=30^{\rm{o}}$ along different $x-z$ planes. The solid line shows the bubble shape along the central axis, while the dashed line and the dotted dashed line show the profile at $y=$100 $\mu$m and $y=$200 $\mu$m off from the centerline, respectively. The inset plot shows a zoom-in profile near the centerline. Figure \ref{fig:x_profile} shows that the thickness of the film increases with increasing $\theta$. In addition, the dashed lines in Fig. \ref{fig:x_profile} indicate that the dimple size is less than 100 $\mu$m from the center along $y$. Figure \ref{fig:y_profile} shows the bubble 2D profile in the $yz$ plane at $x=0$ $\mu$m, 100 $\mu$m, and 200 $\mu$m with the titled angle of $\theta=10^{\rm{o}}$, $20^{\rm{o}}$, and $30^{\rm{o}}$. The bubble profile is symmetric about the plane of $y=0$ as shown in Fig. \ref{fig:y_profile}. Additionally, Fig. \ref{fig:y_profile} shows that the dimple width decreases with increasing $\theta$.

\begin{figure}
\includegraphics[width=\textwidth]{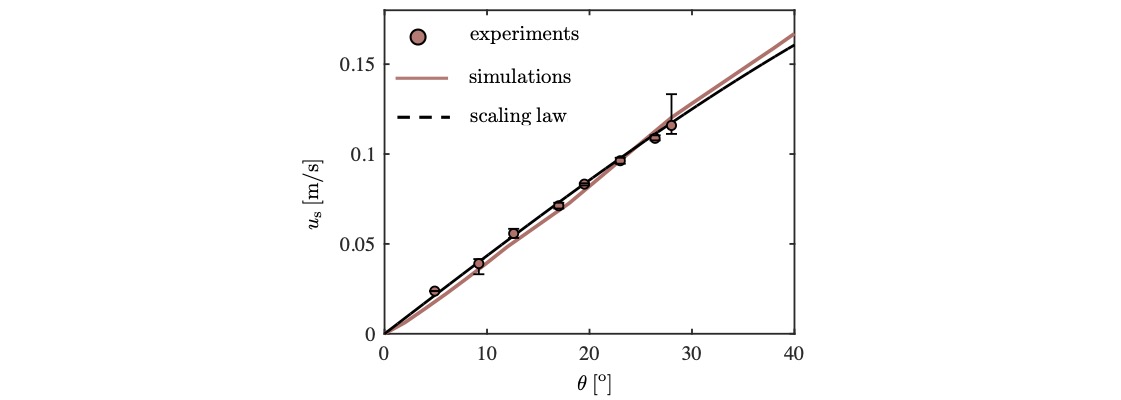}
\caption{\label{fig:u_s} Comparison of the steady-state sliding velocity, $u_{\rm{s}}$ for varying $\theta$ between the experiments (symbols), the simulations (solid line), and the scaling laws (dashed line). The dashed line shows $u_{\rm{s}}=0.25\sin{\theta}$. The errorbars represent 3 trials.  }
\end{figure}

Figure \ref{fig:u_s} shows the steady sliding velocity, $u_{\rm{s}}$, with different angles. As indicated in Fig. \ref{fig:u_s}, $u_{\rm{s}}$ computed from numerical simulations is in good agreement with the experiments. The experimental data presented in Fig. \ref{fig:u_s} are collected in a separate series of experiments in which single air bubbles with $a=0.6$ mm and a standard deviation of 0.03 mm are recorded while injected individually. To discuss our results, we consider a bubble sliding along the surface due to the buoyancy force while experiencing an opposing drag force. Along the direction of sliding, the force balance shows $\rho g \Omega \sin \theta \simeq 6 \pi \mu a u_s$, which leads to  $u_s \simeq (\rho g \Omega/6 \pi \mu a) \sin \theta = (2 \rho g a^2/9 \mu) \sin \theta \approx 0.4 \sin \theta$ [m/s] using $\rho = 980$ kg\,m$^{-3}$, $g=9.8$ m\,s$^{-2}$, $a = 6\times 10^{-4}$ m, and $\mu \approx 2 \times 10^{-3}$ kg\,m$^{-1}$\,s$^{-1}$. This relation is slightly off from the best-fit $u_s = 0.25 \sin \theta$ [m/s] by a factor of 1.6 only (see Fig. \ref{fig:u_s}).

\begin{figure}
\includegraphics[width=\textwidth]{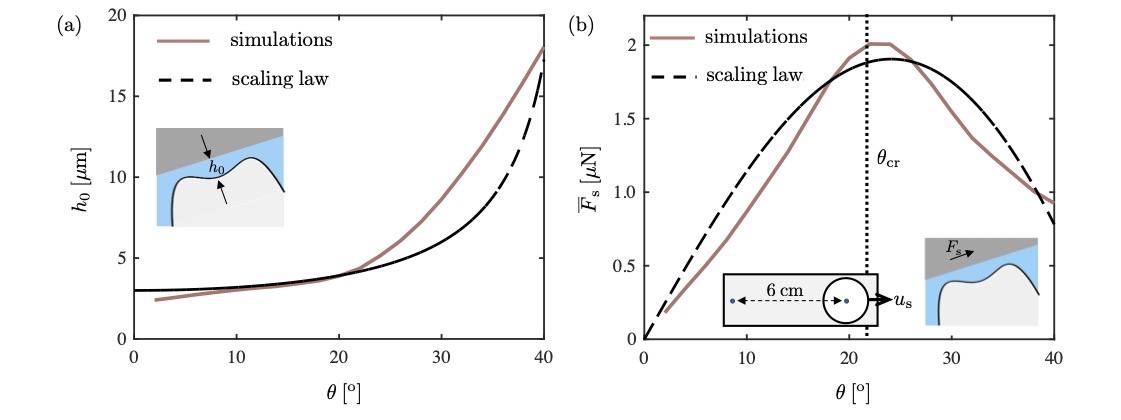}
\caption{\label{fig:F_s} (a) Simulation (solid line), and scaling law (dashed line) results for the steady film thickness in the center of the bubble, $h_0$, at varying $\theta$. The dashed line shows $h_0/cos(2\theta)$ with $h_0=3\: \mu \rm{m}$. (b) Simulation (solid line), and scaling law (dashed line) results for the shear force, $\overline{F}_{\rm{s}}$, over the surface averaged from $X=0-6$ cm at varying $\theta$. The dashed line shows $\overline{F}_{\rm{s}}=7\times 10^{-6} \sin \theta \cos (2\theta)$.  }
\end{figure}

Next, we consider the force balance along the normal direction to the surface. The normal force in the lubrication approximation becomes $\mu u_s (A/h) / \sin \theta$, which balances with the normal buoyancy force $\rho g \Omega \cos \theta$. Under the assumption that the contact area, $A$, is proportional to $a^2$, $h \simeq \frac{\mu A}{\rho g \Omega} \frac{\rho g \Omega}{6 \pi \mu a} (\cos \theta)^{-1} = \frac{A}{6 \pi a} (\cos \theta)^{-1} $. Here, $A$ is the area where the shear force is dominant. We assume that $A$ is the circular domain of radius $=0.2 a$ corresponding to the dimple size. Then, the film thickness becomes $h \simeq 2 (\cos \theta)^{-1} \mu \mathrm{m}$, which shows about 2 $\mu \mathrm{m}$ at $\theta=0$ close to what we found in the simulation in Figs. \ref{fig:x_profile} and \ref{fig:y_profile}. In terms of the $\theta$-dependence, Fig. \ref{fig:F_s} shows that $h$ is proportional to $(\cos 2 \theta)^{-1}$ not $(\cos \theta)^{-1}$. This is presumably due to the complicated interplay between various forces: buoyancy, drag, and thin film force, which cannot not be captured in this simple scaling argument. 

Finally, we compute the mean shear force on the surface along $X$ as $\overline{F}_{\rm{s}}=\int^{6\:\rm{cm}}_{0}F_{\rm{s}} dX$ where  ${F}_{\rm{s}}$ denotes the temporal shear force on the surface. We note that $\overline{F}_{\rm{s}}$ beyond 6 cm does not change significantly. Hence, the resulting average value is close to the steady-state value. However, decreasing the integration domain to a length scale comparable to the length of the bouncing regime may lead to a different result which is out of the scope of the current study. Following the simple analytical expressions $u_s \propto \sin \theta$ and $h \propto (\cos 2 \theta)^{-1}$, the shear stress scales as $\overline{F}_{\rm{s}} \approx \mu u_{\rm{s}} A /h = 0.25\mu \sin \theta  (6 \pi a) \cos (2\theta) \approx 3\pi  \cdot 0.6 \times 10^{-6} \sin \theta \cos (2\theta)~[N]  = 5.65 \sin \theta \cos (2\theta) ~[\mu \rm{N}]$. We then find the roots for the derivative of $\sin \theta \cos (2\theta)$ using Newton's method which gives $\theta_{\rm{cr}}\approx24^{\rm{o}}$ close to the values seen in both experiments and simulations. In addition, since we are in a small slope regime, we can simply add a unity or $\cos \theta$ to the expression for $\overline{F}_{\rm{s}}$. Then, the shear stress becomes $\overline{F}_{\rm{s}} \sim  5.65 \sin \theta \cos \theta \cos (2\theta) ~[\mu \rm{N}] \simeq 2.82 \sin(2 \theta) \cos (2 \theta) ~[\mu \rm{N}] = 1.41 \sin (4 \theta) ~[\mu \rm{N}] $. Therefore, the shear stress is maximized around $\theta \simeq \pi/8$ (i.e., $4\theta =\pi/2$), which is very close to what we observed in both simulations and experiments. It is worth noting that the magnitude of the stress computed in the simulations is a bit higher than the value we predict from the simple theory. This is presumably because our analytical theory estimates the shear stress on a circular dimple and not on the entire bubble.

\section{Discussions}\label{sec:summary}
We examined the cleaning effect of millimetric bubbles at different inclination angles of the substrate coated with either proteins or bacterial biofilm. Our results indicated that both coatings produce similar cleaning results, where $\theta \simeq 22^{\rm{o}} \simeq \pi/8$ gives the highest cleaning parameter, $\lambda$. This consistent result with both protein coating and bacterial coating suggests that surface wettability is not a key factor in cleaning, but that the angle of inclination is. We also computationally investigated the bubble impact at different surface angles from the first impact moment until the bubble reaches the edge of the surface in a steady sliding regime. Our numerical model indicated that while the steady sliding speed increases with the inclination angle, the characteristic film thickness between the bubble and the surface also increases. Since the sliding speed and the film thickness have counter effects on the shear force, their interaction yields the maxima in shear force vs. surface angle.  

The current study investigates the role of surface geometry (i.e., inclination angle) in the use of air bubbles as a sustainable method to effectively sanitize surfaces contaminated with active or passive coatings. This method can be advantageous in cleaning surfaces of soft materials such as fruits and vegetables, where conventional methods damage the soft tissues of the produce. In addition, our findings on the optimal cleaning angle can be leveraged to design bubble-cleaning machinery applicable to biomedical devices. Further studies are required to fully characterize the role of bubble size on the most effective angle to account for polydisperse bubble injection. In addition, studying bubble dynamics on curved surfaces is a relevant topic that will be addressed in the future study.

\section*{Acknowledgments}
This work was supported by the National Science Foundation (NSF) under Grant No. CBET-1919753.

\end{document}